\documentclass[12pt]{article}
\newenvironment{sciabstract}{%
\begin{quote} \bf}
{\end{quote}}

\newcounter{lastnote}

\usepackage{graphicx}
\usepackage{amsmath}
\usepackage{amssymb}
\usepackage{caption}

\title{Fermi-LAT Observations of the Gamma-ray Burst GRB 130427A} 

\author{ 
M.~Ackermann$^{1}$,
M.~Ajello$^{2}$, 
K.~Asano$^{3}$,
W.~B.~Atwood$^{4}$, \\
M.~Axelsson$^{5,6,7}$, 
L.~Baldini$^{8}$,
J.~Ballet$^{9}$,
G.~Barbiellini$^{10,11}$,\\
M.~G.~Baring$^{12}$, 
D.~Bastieri$^{13,14}$
% ...
\and 
\and
\normalsize{1. Deutsches Elektronen Synchrotron DESY, D-15738 Zeuthen, Germany}\\ 
\normalsize{2. Space Sciences Laboratory, 7 Gauss Way, University of California,}\\ 
\normalsize{Berkeley, CA 94720-7450, USA}\\ 
\normalsize{3. ICRR}\\ 
\normalsize{4. Santa Cruz Institute for Particle Physics, Department of Physics }\\ 
\normalsize{and Department of Astronomy and Astrophysics, University of California }\\ 
\normalsize{at Santa Cruz, Santa Cruz, CA 95064, USA}\\ 
\normalsize{5. Department of Astronomy, Stockholm University, SE-106 91 Stockholm, }\\ 
\normalsize{Sweden}\\ 
\normalsize{6. The Oskar Klein Centre for Cosmoparticle Physics, AlbaNova, SE-106 }\\ 
\normalsize{91 Stockholm, Sweden}\\ 
\normalsize{7. Department of Physics, Royal Institute of Technology (KTH), AlbaNova,}\\ 
\normalsize{ SE-106 91 Stockholm, Sweden}\\ 
\normalsize{8. Universit\`a  di Pisa and Istituto Nazionale di Fisica Nucleare, }\\ 
\normalsize{Sezione di Pisa I-56127 Pisa, Italy}\\ 
\normalsize{9. Laboratoire AIM, CEA-IRFU/CNRS/Universit\'e Paris Diderot, Service }\\ 
\normalsize{d'Astrophysique, CEA Saclay, 91191 Gif sur Yvette, France}\\ 
\normalsize{10. Istituto Nazionale di Fisica Nucleare, Sezione di Trieste, I-34127}\\ 
\normalsize{ Trieste, Italy}\\ 
\normalsize{11. Dipartimento di Fisica, Universit\`a di Trieste, I-34127 Trieste,}\\ 
\normalsize{ Italy}\\ 
\normalsize{12. Rice University, Department of Physics and Astronomy, MS-108, }\\ 
\normalsize{P. O. Box 1892, Houston, TX 77251, USA}\\ 
\normalsize{13. Istituto Nazionale di Fisica Nucleare, Sezione di Padova, I-35131 }\\ 
\normalsize{Padova, Italy}\\ 
\normalsize{14. Dipartimento di Fisica e Astronomia ``G. Galilei'', Universit\`a }\\ 
\normalsize{di Padova, I-35131 Padova, Italy}\\ 
\normalsize{15. W. W. Hansen Experimental Physics Laboratory, Kavli Institute for }\\ 
\normalsize{Particle Astrophysics and Cosmology, Department of Physics and SLAC }\\ 
\normalsize{National Accelerator Laboratory, Stanford University, Stanford, CA }\\ 
\normalsize{94305, USA}\\ 
\normalsize{16. Istituto Nazionale di Fisica Nucleare, Sezione di Pisa, I-56127 }\\ 
\normalsize{Pisa, Italy}\\ 
\and
\normalsize{17. Istituto Nazionale di Fisica Nucleare, Sezione di Trieste, and }\\ 
\normalsize{Universit\`a di Trieste, I-34127 Trieste, Italy}\\ 
\normalsize{18. Istituto Nazionale di Fisica Nucleare, Sezione di Perugia, }\\ 
\normalsize{I-06123 Perugia, Italy}\\ 
\normalsize{19. Dipartimento di Fisica, Universit\`a degli Studi di Perugia, }\\ 
\normalsize{I-06123 Perugia, Italy}\\ 
\normalsize{20. Dipartimento di Fisica ``M. Merlin" dell'Universit\`a e del }\\ 
\normalsize{Politecnico di Bari, I-70126 Bari, Italy}\\ 
\normalsize{21. Istituto Nazionale di Fisica Nucleare, Sezione di Bari, 70126 Bari,}\\ 
\normalsize{ Italy}\\ 
\normalsize{22. Laboratoire Leprince-Ringuet, \'Ecole polytechnique, CNRS/IN2P3, }\\ 
\normalsize{Palaiseau, France}\\ 
\normalsize{23. Center for Space Plasma and Aeronomic Research (CSPAR), University }\\ 
\normalsize{of Alabama in Huntsville, Huntsville, AL 35899, USA}\\ 
\normalsize{24. Institut de Ci\`encies de l'Espai (IEEE-CSIC), Campus UAB, 08193 }\\ 
\normalsize{Barcelona, Spain}\\ 
\normalsize{25. INAF-Istituto di Astrofisica Spaziale e Fisica Cosmica, I-20133 }\\ 
\normalsize{Milano, Italy}\\ 
\normalsize{26. Center for Earth Observing and Space Research, College of Science,}\\ 
\normalsize{ George Mason University, Fairfax, VA 22030, resident at Naval Research }\\ 
\normalsize{ Laboratory, Washington, DC 20375, USA}\\ 
\normalsize{27. Space Science Division, Naval Research Laboratory, Washington, }\\ 
\normalsize{DC 20375-5352, USA}\\ 
\normalsize{28. Agenzia Spaziale Italiana (ASI) Science Data Center, I-00044 }\\ 
\normalsize{Frascati (Roma), Italy}\\ 
\normalsize{29. Istituto Nazionale di Astrofisica - Osservatorio Astronomico }\\ 
\normalsize{di Roma, I-00040 Monte Porzio Catone (Roma), Italy}\\ 
\normalsize{30. Universities Space Research Association (USRA), Columbia, }\\ 
\normalsize{MD 21044, USA}\\ 
\normalsize{31. Laboratoire Univers et Particules de Montpellier, Universit\'e }\\ 
\normalsize{Montpellier 2, CNRS/IN2P3, Montpellier, France}\\ 
\normalsize{32. NASA Postdoctoral Program Fellow, USA}\\ 
\normalsize{33. Department of Physics and Astronomy, Sonoma State University, }\\ 
\normalsize{Rohnert Park, CA 94928-3609, USA}\\ 
\and
\normalsize{34. Department of Physics, Stockholm University, AlbaNova, SE-106 91 }\\ 
\normalsize{Stockholm, Sweden}\\ 
\normalsize{35. Royal Swedish Academy of Sciences Research Fellow, funded by a grant }\\ 
\normalsize{from the K. A. Wallenberg Foundation}\\ 
\normalsize{36. The Royal Swedish Academy of Sciences, Box 50005, SE-104 05 }\\ 
\normalsize{Stockholm, Sweden}\\ 
\normalsize{37. INAF Istituto di Radioastronomia, 40129 Bologna, Italy}\\ 
\normalsize{38. Dipartimento di Fisica, Universit\`a di Udine and Istituto Nazionale}\\ 
\normalsize{ di Fisica Nucleare, Sezione di Trieste, Gruppo Collegato di Udine, }\\ 
\normalsize{ I-33100 Udine, Italy}\\ 
\normalsize{39. Stellar Solutions Inc., 250 Cambridge Avenue, Suite 204, Palo Alto,}\\ 
\normalsize{ CA 94306, USA}\\ 
\normalsize{40. Jacobs Technology, Huntsville, AL 35806, USA}\\ 
\normalsize{41. NASA Goddard Space Flight Center, Greenbelt, MD 20771, USA}\\ 
\normalsize{42. University College Dublin, Belfield, Dublin 4, Ireland}\\ 
\normalsize{43. Department of Physical Sciences, Hiroshima University, }\\ 
\normalsize{Higashi-Hiroshima, Hiroshima 739-8526, Japan}\\ 
\normalsize{44. Department of Natural Sciences, The Open University of Israel, }\\ 
\normalsize{1 University Road, POB 808, Ra'anana 43537, Israel}\\ 
\normalsize{45. Max-Planck Institut f\"ur extraterrestrische Physik, 85748 Garching, }\\ 
\normalsize{Germany}\\ 
\normalsize{46. Department of Astronomy, Graduate School of Science, Kyoto University, }\\ 
\normalsize{Sakyo-ku, Kyoto 606-8502, Japan}\\ 
\normalsize{47. Department of Physics, Center for Cosmology and Astro-Particle Physics,}\\ 
\normalsize{ The Ohio State University, Columbus, OH 43210, USA}\\ 
\normalsize{48. Science Institute, University of Iceland, IS-107 Reykjavik, Iceland}\\ 
\normalsize{49. CNRS, IRAP, F-31028 Toulouse cedex 4, France}\\ 
\normalsize{50. GAHEC, Universit\'e de Toulouse, UPS-OMP, IRAP, Toulouse, France}\\ 
\normalsize{51. Istituto Nazionale di Fisica Nucleare, Sezione di Torino, I-10125 }\\ 
\normalsize{Torino, Italy}\\ 
\normalsize{52. Department of Physics and Department of Astronomy, University of }\\ 
\normalsize{Maryland, College Park, MD 20742, USA}\\ 
\and
\normalsize{53. Hiroshima Astrophysical Science Center, Hiroshima University, }\\ 
\normalsize{Higashi-Hiroshima, Hiroshima 739-8526, Japan}\\ 
\normalsize{54. Center for Research and Exploration in Space Science and Technology}\\ 
\normalsize{ (CRESST) and NASA Goddard Space Flight Center, Greenbelt, MD 20771, USA}\\ 
\normalsize{55. Istituto Nazionale di Fisica Nucleare, Sezione di Roma ``Tor Vergata",}\\ 
\normalsize{ I-00133 Roma, Italy}\\ 
\normalsize{56. Solar-Terrestrial Environment Laboratory, Nagoya University, }\\ 
\normalsize{Nagoya 464-8601, Japan}\\ 
\normalsize{57. Max-Planck-Institut f\"ur Physik, D-80805 M\"unchen, Germany}\\ 
\normalsize{58. Department of Physics and Center for Space Sciences and Technology, }\\ 
\normalsize{University of Maryland Baltimore County, Baltimore, MD 21250, USA}\\ 
\normalsize{59. University of Johannesburg, Department of Physics, University of }\\ 
\normalsize{Johannesburg, Auckland Park 2006, South Africa, }\\ 
\normalsize{60. Institut f\"ur Astro- und Teilchenphysik and Institut f\"ur }\\ 
\normalsize{Theoretische Physik, Leopold-Franzens-Universit\"at Innsbruck, A-6020 }\\ 
\normalsize{Innsbruck, Austria}\\ 
\normalsize{61. Department of Physics, University of Washington, Seattle, WA 98195-1560,}\\ 
\normalsize{ USA}\\ 
\normalsize{62. Space Sciences Division, NASA Ames Research Center, Moffett Field, }\\ 
\normalsize{CA 94035-1000, USA}\\ 
\normalsize{63. NYCB Real-Time Computing Inc., Lattingtown, NY 11560-1025, USA}\\ 
\normalsize{64. Ad{\i}yaman University, 02040 Ad{\i}yaman, Turkey}\\ 
\normalsize{65. Instituci\'o Catalana de Recerca i Estudis Avan\c{c}ats (ICREA), }\\ 
\normalsize{Barcelona, Spain}\\ 
\normalsize{66. Consorzio Interuniversitario per la Fisica Spaziale (CIFS), }\\ 
\normalsize{I-10133 Torino, Italy}\\ 
\normalsize{67. Dipartimento di Fisica, Universit\`a di Roma ``Tor Vergata", }\\ 
\normalsize{I-00133 Roma, Italy}\\ 
\normalsize{68. Department of Physics and Mathematics, Aoyama Gakuin University, }\\ 
\normalsize{Sagamihara, Kanagawa, 252-5258, Japan}\\ 
\normalsize{69. NASA Marshall Space Flight Center, Huntsville, AL 35812, USA}\\ 
\normalsize{70. Los Alamos National Laboratory, Los Alamos, NM 87545, USA}\\ 
\normalsize{71. Astronomical Institute "Anton Pannekoek" University of Amsterdam, }\\ 
\normalsize{Postbus 94249 1090 GE Amsterdam, Netherlands}\\ 
\and
\normalsize{72. INAF - Istituto di Astrofisica Spaziale e Fisica Cosmica, }\\ 
\normalsize{Via U. La Malfa 153, I-90146 Palermo, Italy}\\ 
\normalsize{73. University of Maryland, Baltimore County / Center for Research and }\\ 
\normalsize{Exploration in Space Science and Technology}\\ 
\normalsize{}\\ 
\normalsize{$^\ast$To whom correspondence should be addressed. Email: }\\ 
\and
\normalsize{sjzhu@umd.du (S.J.Z.); jchiang@slac.stanford.edu (J.C.);}\\ 
\normalsize{charles.dermer@nrl.navy.mil (C.D.D.); nicola.omodei@stanford.edu (N.O.); }\\ 
\normalsize{giacomov@slac.stanford.edu (G.V.); Shaolin.Xiong@uah.edu (S.X.).}\\ 
}

\date{}

\begin{document}
\maketitle 

\begin{sciabstract}

The observations of the exceptionally bright gamma-ray burst (GRB) 130427A by the Large Area Telescope aboard the Fermi Gamma-ray Space Telescope provide constraints on the nature of such unique astrophysical sources. GRB 130427A had the largest fluence, highest-energy photon (95~GeV), longest $\gamma$-ray duration (20 hours), and one of the largest isotropic energy releases ever observed from a GRB. Temporal and spectral analyses of GRB 130427A challenge the widely accepted model that the non-thermal high-energy emission in the afterglow phase of GRBs is synchrotron emission radiated by electrons accelerated at an external shock.
\end{sciabstract}

\section*{Introduction}

Gamma-ray bursts (GRBs) are thought to originate from collapsing massive stars or merging compact objects (such as neutron stars or black holes), and are associated with the formation of black holes in distant galaxies.

GRB 130427A was detected by both the Large Area Telescope (LAT) and the Gamma-ray Burst Monitor (GBM) aboard the Fermi Gamma-ray Space Telescope. The LAT is a pair-conversion telescope that observes photons from 20~MeV to $>\!\!$~300 GeV with a 2.4~steradian field of view ({\it 1}). The GBM consists of 12 sodium iodide (NaI, 8~keV -- 1~MeV) and 2 bismuth germanate (BGO, 200~keV -- 40~MeV) detectors, positioned around the spacecraft to view the entire unocculted sky ({\it 2}).

In the standard model of GRBs, the blast wave that produces the initial, bright prompt emission later collides with the external material surrounding the GRB (the circumburst medium) and creates shocks (see, e.g., ({\it 3})). These external shocks accelerate charged particles, which produce photons through synchrotron radiation. Until this burst, the high-energy emission from LAT-detected GRBs had been well described by this model, but GRB 130427A challenges this widely accepted model. In particular, the maximum possible photon energy prescribed by this model is surpassed by the the late-time high-energy photons detected by the LAT. The LAT detected high-energy $\gamma$-ray emission from this burst for almost a day, including a 95~GeV photon (which was emitted at 128~GeV in the rest frame at redshift $z = 0.34$ ({\it 4})) a few minutes after the burst began and a 32~GeV photon (43~GeV in the rest frame) after more than 9 hours. These are more energetic and detected at considerably later times than the previous record holder, an 18~GeV photon detected by the Energetic Gamma Ray Experiment Telescope (EGRET) aboard the  Compton Gamma Ray Observatory more than 90~minutes after GRB 940217 began ({\it 5}).

\section*{Observations}

At 07:47:06.42 UTC on 27 April 2013 ($T_0$), while Fermi was in the regular survey mode, the GBM triggered on GRB 130427A. The burst was sufficiently hard and intense to initiate an Autonomous Repoint Request ({\it 6}), a spacecraft slewing maneuver that keeps the burst within the LAT field of view for 2.5 hours, barring Earth occultation. At the time of the GBM trigger, the Swift Burst Alert Telescope (BAT) was slewing between two pre-planned targets, and triggered on the ongoing burst at 07:47:57.51 UTC immediately after the slew completed ({\it 7}), 51.1~s after the GBM trigger. The CARMA millimeter-wave observatory localized this burst to (R.A., Dec.) = (173.1367$^\circ$, 27.6989$^\circ$) (J2000) with a $0.4''$ uncertainty ({\it 8}). The Rapid Telescopes for Optical Response (RAPTOR) detected bright optical emission from the GRB, peaking at a red-band magnitude of R = 7.03 $\pm$ 0.03 around the GBM trigger time before fading to R$\sim$10 about 80 seconds later ({\it 9}). The Gemini-North observatory reported a redshift of $z=0.34$ ({\it 4}), and an underlying supernova has been detected ({\it 10}). A total of 58 observatories have reported observations of this burst as of September 2013.

At the time of the GBM trigger, the GRB was 47.3$^\circ$ from the LAT boresight, well within the LAT field of view. The Autonomous Repoint Request brought the burst to 20.1$^\circ$ from the LAT boresight based on the position calculated by the GBM flight software. It remained in the LAT field of view for 715~s until it became occulted by the Earth, re-emerging from Earth occultation at $T_0$~+~3135~s. Within the first $\sim$80~ks after the trigger, the LAT detected more than 500 photons with energies $>$100~MeV associated with the GRB; the previous record holder was GRB 090902B, with $\sim$200 photons ({\it 11}). In addition, the LAT detected 15 photons with energies $>$10~GeV (compared to only 3 photons for GRB 090902B).  Using the LAT Low Energy (LLE) event selection ({\it 12}), which considerably increases the LAT effective collecting area to lower-energy $\gamma$-rays down to 10~MeV (with adequate energy reconstruction down to 30~MeV) ({\it 13, 14}), thousands of counts above background were detected between $T_0$ and $T_0+100$~s.

\section*{Temporal characteristics}
The temporal profile of the emission from GRB 130427A varies strongly with energy from 10~keV to $\sim\!\! 100$~GeV (Fig.~1 and Fig.~2). The GBM light curves consist of an initial peak with a few-second duration, a much brighter multipeaked emission episode lasting $\sim\!\! 10$~s, and a dim, broad peak at $T_0 + 120$~s, which fades to an undetectable level after $\sim\! 300$~s (also seen in the Swift light curve ({\it 7})).

The triggering pulse observed in the LLE ($>$10~MeV) light curve is more sharply peaked than the NaI- and BGO-detected emission at $T_0$. The LLE light curve between $T_0+4$~s and $T_0+12$~s exhibits a multipeaked structure. Some of these peaks have counterparts in the GBM energy range, although the emission episodes are not perfectly correlated (e.g., the sharp spike in the LLE light curve at $T_0 + 9.5$~s is not relatively bright in the GBM light curves) because of the spectral evolution with energy.

The LAT-detected emission, however, does not appear to be temporally correlated with either the LLE or GBM emission beyond the initial spike at $T_0$. Photons with energies $>$1~GeV are first observed $\sim$10 s after $T_0$, after the brightest GBM emission has ended, consistent with a delayed onset of the high-energy emission ({\it 13}). The delayed onset is not caused by a progressively increasing LAT acceptance due to slewing, because the slew started at $T_0+33$~s. Instead, it reflects the true evolution of the GRB emission.

GRB spectra are generally well described by phenomenological models such as the Band function ({\it 15}) or the smoothly broken power law (SBPL ({\it 16})). For the brightest LAT bursts, the onset of the GeV emission is delayed with respect to the keV-MeV emission and can be fit by an additional power-law component ({\it 13}). This additional component usually becomes significant while the keV-MeV emission is still bright.

For GRB 130427A, however, the extra power-law component becomes significant only after the GBM-detected emission has faded (Fig.~3). During the initial peak ($T_0-0.1$~s to $T_0+4.5$~s), there are only a few LAT-detected photons, and the emission is well-fit by an SBPL. For the brightest part of the burst ($T_0+4.5$~s to $T_0+11.5$~s), we did not use the GBM-detected emission because of the substantial systematic effects caused by extremely high flux (SOM); however, there are no photons with energies greater than 1~GeV in this time interval, and the energy spectrum $>\!\!30$~MeV is well described by a single power law without a break (Fig.~3) (note that the LAT did not suffer from any pile up issues). Photons with energies greater than 1~GeV are detected in the last time interval ($T_0+11.5$~s to $T_0+33.0$~s), including a 73~GeV photon at $T_0+19$~s. Unlike other bright LAT bursts, the LAT-detected emission from GRB 130427A appears to be temporally distinct from the GBM-detected emission, suggesting that the GeV and keV-MeV photons arise from different emission regions or mechanisms.

\section*{Temporally extended high-energy emission}

To characterize the temporally extended high-energy emission, we performed an unbinned maximum likelihood analysis of the LAT $>$100 MeV data. We modeled the LAT photon spectrum as a power law with a spectral index $\alpha$ (i.e., the spectrum $N(E) \propto E^\alpha$). We found evidence of spectral evolution during the high-energy emission. In contrast to  another study ({\it 17}),  that used longer time intervals in the spectral fits, we found that the LAT $>$100~MeV spectrum of the GRB is well described by a power law at all times, but with a varying spectral index (SOM).

Fig.~2 shows the behavior of the temporally extended LAT emission. During the first pulse around $T_0$, the $>$100 MeV emission is faint and soft; the pulse contains only a few photons, and their energies are all $<$1 GeV. This is followed by a period during which there is no significant $>$100 MeV emission, while the GBM emission is at its brightest. Starting at $\sim \!\!T_0+5$~s, the $>$100 MeV emission is detectable again, but remains dim until $\sim \!\!T_0+12$~s. The spectral index fluctuates between $\alpha \sim -2.5$ and $\alpha \sim -1.7$. At late times ($>\!T_0 + 300$~s), we measured typical spectral indices of $\alpha \sim -2$, consistent with the indices of other LAT bursts ({\it 13}). During the time intervals with the hardest spectra, the LAT observed the highest energy photons --- such as the 73 GeV photon at $T_0+19$~s and the record breaking 95 GeV photon at $T_0+244$~s --- which severely restrict the possible mechanisms that could generate the high-energy afterglow emission (Table S2).

The temporally extended photon flux light curve is better fit by a broken power law than a power law. We found a break after a few hundred seconds, with the temporal index steepening from $-0.85 \pm 0.08$ to $-1.35 \pm 0.08$ ($\chi^2$/dof = 36/19 for a single power law, 16/17 for a broken power law). In contrast, a break is not statistically preferred in the energy flux light curve ($\chi^2$/dof = 14/18 for a single power law, 13/17 for a broken power law), probably because of the larger statistical uncertainties. For a single power-law fit to the energy flux light curve, we found a temporal index of $-1.17 \pm 0.06$, consistent with other LAT bursts ({\it 13}).

The GBM and Swift energy flux light curves are also shown in Fig.~2. The Swift X-Ray Telescope (XRT) began observing the burst at $T_0+190$~s; the reported XRT+BAT (0.3 -- 10 keV) light curve is a combination of XRT data and BAT-detected emission (15 -- 150 keV) extrapolated down into the energy range of the XRT. The XRT+BAT light curve shows the unabsorbed flux in the 0.3 -- 10~keV range ({\it 7}). During the initial part of the burst, the GBM (10~keV -- 10~MeV) light curve peaks earlier than both the XRT+BAT (0.3 -- 10~keV) and LAT ($>$100~MeV) light curves. The GBM light curve peaks again at $\sim\!T_0+120$~s (see also ({\it 7})), while the LAT light curve shows a sharp and hard peak at $T_0+200$~s. The BAT+XRT light curve peaks again as well at the same time as the LAT light curve, but the peak is much broader.

\section*{Interpretation}
The energetics of GRB 130427A place it among the brightest LAT bursts. For GRB 130427A, the 10 keV -- 20~MeV fluence measured with the GBM in the 400 s following $T_0$ is
$\sim\!4.2 \times 10^{-3}$~erg\,cm$^{-2}$. The issue with pulse pileup and the uncertainties in the calibration of the GBM detectors contribute to a systematic error which we estimate to be less than 20\%; the statistical uncertainty ($0.01 \times 10^{-3}$ erg cm$^{-2}$ is negligible with respect to the systematic one (SOM). The $>$100~MeV fluence measured with the LAT in the 100~ks following $T_0$ is $(7 \pm 1)\times 10^{-4}$~erg\,cm$^{-2}$.  The total LAT fluence is therefore $\approx\!20$\% of the GBM fluence, similar to other bright LAT GRBs ({\it 13, 19}). For a total 10~keV -- 100~GeV fluence of $4.9\times 10^{-3}$~erg\,cm$^{-2}$, the total apparent isotropic $\gamma$-ray energy (i.e., the total energy release if there were no beaming) is ${\cal E}_{\gamma,iso} = 1.40\times 10^{54}$~erg, using a flat $\Lambda$CDM cosmology with $h = 0.71$ and $\Omega_\Lambda = 0.73$, implying a luminosity distance of 1.8~Gpc for $z = 0.34$. This value of ${\cal E}_{\gamma,iso}$ is only slightly less than the values for other bright LAT hyper-energetic events, which include GRB 080916C, GRB 090902B, and
GRB 090926A ({\it 19})

The emission region must be transparent against absorption by photon-photon pair production, which has a significant effect at the energies of the LAT-detected emission. Viable models of
GRBs therefore require highly relativistic, jetted plasma outflows with bulk jet Lorentz factors $\Gamma \gtrsim 100$ ({\it 20}). The 73~GeV photon at $T_0 + 19$~s (Table S2) provides the most stringent limit on $\Gamma$. Assuming that the variability timescale reflects the size of the emitting region, and that the MeV and GeV emissions around the time of the 73~GeV photon at $T_0 + 19$~s are cospatial, the requirement that the optical depth due to $\gamma\gamma$ opacity be less than 1 then implies that the minimum bulk Lorentz factor is $\Gamma_{min} = 455^{+16}_{-13}$. Here a SBPL fit to the GBM spectrum in the 11.5 -- 33.0~s interval (Table S1) and a minimum variability timescale of $0.04 \pm 0.01$~s are used (SOM). The cospatial assumption is, however, questionable given the different time histories in the MeV and GeV emission. Moreover, values of $\Gamma_{min}$ that are smaller by a factor of 2 -- 3 can be realized for models with time-dependent $\gamma$-ray opacity in a thin-shell model ({\it 21}).

The delayed onset of the LAT-detected emission with respect to the GBM-detected emission is an important clue to the nature of GRBs ({\it 13}). For GRB 130427A, the LAT-detected emission becomes harder and more intense after the GBM-detected emission has faded (Fig.~3). This suggests that the GeV emission is produced later than the keV-MeV emission and in a different region. In particular, if the keV-MeV emission comes from interactions within the outflow itself, the GeV emission arises from the outflow's interactions with the circumburst medium.

The explosive relativistic outflow of a GRB sweeps up and drives a shock into the circumburst medium. The medium could have, for instance, a uniform density $n_0$ (cm$^{-3}$) or a $n(r)\propto r^{-2}$ density profile resulting from the stellar wind of the Type Ic supernova progenitor star associated with GRB 130427A ({\it 10}). The LAT observations of GRB 130427A challenge the scenario in which the GeV photons are nonthermal synchrotron radiation emitted by electrons accelerated at the external forward shock ({\it 22, 23}). In this model, the time of the brightest emission corresponds to the time $t_{dec}$ when most of the outflow energy is transferred to the shocked external medium.  The Lorentz factor $\Gamma(t_{dec})$ of the shock outflow in the external medium at the deceleration time $t_{dec}$ is a lower limit for the initial bulk outflow Lorentz factor $\Gamma_0$, because a relativistic reverse shock would lower the shocked fluid Lorentz factor below $\Gamma_0$, and the engine timescale $T_{GRB}$ can be longer than $t_d$, the deceleration timescale for an impulsive explosion.  The activity of the central engine that produces the blast wave is revealed by the keV -- MeV emission from particles accelerated at colliding-wind shocks. For GRB 130427A, the GeV emission starts to decay as a power law in time by $t\approx 20$~s (Fig.~2), and most of the keV-MeV radiation has subsided by $t\approx 12$~s (Fig.~1).  The blast wave is in the self-similar deceleration phase at $t > t_{dec} = \max [t_{d},T_{GRB}]$, where $T_{GRB}$ is the engine timescale (over which most of the outflow energy was released). Here  $t_{d} ({\rm s}) \cong 2.4 [({\cal E}_{\gamma,iso}/10^{54}{\rm erg})/n_0]^{1/3}/(\Gamma_0/1000)^{8/3}\;$ for a uniform external medium, and $t_{d} ({\rm s}) \cong 6.3 (\mathcal{E}_{\gamma,iso}/10^{55} \rm erg) (0.1/A_*)(500/\Gamma_0)^4$ for a stellar wind medium of density $\rho = AR^{-2}$, with $A = 5\times 10^{11}A_*$ g cm$^{-1}$.

Most of the fluence from GRB 130427A was radiated before $t \approx 12\;$s, suggesting that $t_{d} \lesssim 12-15$~s. Defining $t_1 = t_{d}/$(10~s) yields $t_1 \approx 1 - 2$, which gives $\Gamma(t_{dec}) \cong 540 [{\cal E}_{55}/t_1^3 n_0({\rm~cm}^{-3})]^{1/8}$ for the uniform density case, where ${\cal E}_{55} = {\cal E}_{\gamma,iso}/(10^{55}{\rm~erg})$ is the isotropic energy release of the GRB. For a wind medium, $\Gamma(t_{dec}) \cong 450\{\mathcal{E}_{55}/[(A_*/0.1)t_1]\}^{1/4}.$ Both values are close to the $\gamma\gamma$ opacity estimate of $\Gamma_{min}$.
 
The presence of high-energy photons at times $t\gg t_{dec}$ (Table S2) is incompatible with these $\gamma$ rays having a synchrotron origin. Equating the electron energy-loss time scale due to synchrotron radiation with the Larmor timescale for an electron to execute a gyration gives a conservative limit on the maximum synchrotron photon energy $E_{max,syn}\approx {2^{3/2}(27/16\pi\alpha_f)} m_ec^2 \Gamma(t)/(1+z) \approx 79 \Gamma(t)$~MeV (where $\alpha_f$ is the fine structure constant (SOM)).  Using $\Gamma(t)$ derived by Blandford \& McKee
({\it 24}) in the adiabatic limit, we find that the maximum synchrotron photon energy $E_{max,syn}\ll 7 ({\cal E}_{55}/n_0)^{1/8} [t/200{\rm~s}]^{-3/8} $~GeV, which agrees with results from integration over surfaces of equal arrival time in  the self-similar regime ({\it 25}), when a scaling factor of $27/16\pi$ is included (SOM). The presence of a 95~GeV photon at $T_0+ 244$~s (Fig.~4 and Table S2) is incompatible with a synchrotron origin even for conservative assumptions about Fermi acceleration. This conclusion holds for adiabatic and radiative external shocks in both uniform or wind media (see also ({\it 26})). The question of a wind or uniform density model is not settled, but combined forward and reverse shock blast-wave model fits to the radio through X-ray emission from 0.67~d to 9.7~d after the GRB favor a wind medium ({\it 27}), whereas inferences from Swift and LAT data suggest a uniform environment around GRB 130427A ({\it 7}). Even in the extreme case where acceleration is assumed to operate on a timescale shorter than the Larmor timescale by a factor of $2\pi$, synchrotron radiation cannot account for the presence of high-energy
radiation in the afterglow. Synchrotron emission above $\gtrsim 100$ GeV is still possible, however, if an acceleration mechanism faster than the Fermi process is acting, such as magnetic reconnection (e.g., ({\it 28})).

The 95~GeV photon in the early afterglow and the 32~GeV photon at $T_0+34.4$~ks therefore cannot originate from lepton synchrotron radiation in the standard afterglow model with shock Fermi
acceleration (Fig.~4). If the emission mechanism for the GeV photons is not synchrotron radiation, the highest energy photons can still be produced by lepton Compton processes. Synchrotron
self-Compton (SSC) $\gamma$-rays, made when target synchrotron photons are Compton scattered by the same jet electrons that emit the synchrotron emission, is unavoidable. SSC emission is expected to peak at TeV and higher energies during the prompt phase (although no GRB has been observed at TeV energies), and would cause the GeV light curve to flatten and the LAT spectrum to harden when the peak of the SSC component passes through the LAT waveband ({\it 29-31}). Such a feature may be seen in the light curves of GRB 090902B and GRB 090926A at 15~s -- 30~s after
$T_0$ ({\it 13}), but no such hardening or plateau associated with the SSC component is observed in the LAT light curve of GRB 130427A, though extreme parameters might still allow an SSC interpretation; see, e.g., ({\it 32}). Except for the hard flare at $t \approx 250$~s and a possible softening at $\sim 3000$~s (and therefore not associated with a probable beaming break at $t \approx 0.8$~d ({\it 7})), neither the integral photon or energy-flux light curves in Fig.~2 show much structure or strong evidence for temporal or spectral variability from $t= 20$~s to $t=1$~d. The NuSTAR observations of the late-time hard x-ray afterglow provide additional evidence for a single spectral component ({\it 33}).

These considerations suggest that other extreme high-energy radiation mechanisms may be operative, such as external Compton processes. The most intense source of target photons is the powerful engine emissions, as revealed by the GBM and XRT prompt emission. A cocoon or remnant shell is also a possible source of soft photons, but unless the target photon source is extended and
radiant, it would be difficult to model the nearly structureless LAT light curve over a long period of time. Given the similarity between the XRT and LAT light curves Fig.~2), afterglow synchrotron radiation made by electrons accelerated at an external shock would also be the favored explanation for the LAT emission, but this is inconsistent with the detection of high-energy photons at late time.

The photon index of GRB 130427A, $\approx -2$, is similar to those found in calculations of electromagnetic cascades created when the $\gamma$-ray opacity of ultra-high energy (UHE, $>$100 TeV) 
photons in the jet plasma is large ({\it 34}). An electromagnetic cascade induced by ultra-relativistic hadrons would be confirmed by coincident detection of neutrinos, but even for GRB 130427A with its extraordinary fluence, only a marginal detection of neutrinos is expected with IceCube, and none has been reported ({\it 35}). Because the UHE $\gamma$-ray photons induce cascades both inside the radiating plasma and when they travel through intergalactic space (e.g., ({\it 36, 37}), a leptonic or hadronic cascade component in GRBs, which is a natural extension of
colliding shell and blast wave models, might be required to explain the high-energy emission of GRB 130427A provided that the required energies are not excessive. The observations described in this paper demonstrate non-synchrotron emission in the afterglow phase of the bright GRB 130427A, contrary to the hitherto standard model of GRB afterglows. \newline

\section*{References and Notes}
\begin{enumerate}
\item W. Atwood {\it et al.}, {\it Astrophys. J.} {\bf 697}, 1071 (2009).
\item C. Meegan {\it et al.}, {\it Astrophys. J.} {\bf 702}, 791 (2009).
\item N. Gehrels, S. Razzaque, {\it Front. Phys.} (2013).
\item A. Levan {\it et al.}, {\it GCN Circ. 14455} (2013).
\item K. Hurley {\it et al.}, {\it Nature} {\bf 372}, 652 (1994).
\item A. von Kienlin, {\it GCN Circ. 14473} (2013).
\item A. Maselli {\it et al.}, {\it Science} this edition (2013).
\item D. Perley, {\it GCN Circ. 14494} (2013).
\item W. T. Vestrand {\it et al.}, {\it Science} this edition (2013).
\item D. Xu {\it et al.}, {\it Astrophys. J.} in press; preprint available at http://arxiv.org/abs/1305.6832 (2013).
\item A. Abdo {\it et al.}, {\it Astrophys. J.} {\bf 706}, L138 (2009).
\item LLE is a set of cuts that provides increased statistics at the cost of a higher background level and greatly reduced energy and angular reconstruction accuracy.
\item M. Ackermann {\it et al.}, http://arxiv.org/abs/1303.2908 (2013).
\item V. Pelassa {\it et al.}, http://arxiv.org/abs/1002.2617 (2010).
\item D. Band {\it et al.}, {\it Astrophys. J.} {\bf 413}, 281 (1993).
\item A. Goldstein {\it et al.}, {\it Astrophys. J. Suppl. Ser.} {\bf 199}, 19 (2012).
\item P.-H. T. Tam {\it et al.}, {\it Astrophys. J.} {\bf 771}, L13 (2013).
\item D. Eichler {\it et al.}, {\it Astrophys. J.} {\bf 722}, 543 (2010).
\item S. B. Cenko {\it et al.}, {\it Astrophys. J.} {\bf 732}, 29 (2011).
\item P. Guilbert {\it et al.}, {\it Mon. Not. R. Astron. Soc.} {\bf 205}, 593 (1983).
\item J. Granot {\it et al.}, {\it Astrophys. J.} {\bf 677}, 92 (2008).
\item P. Kumar, R. Barniol-Duran, {\it Mon. Not. R. Astron. Soc.} {\bf 400}, L75 (2009).
\item G. Ghisellini {\it et al.}, {\it Mon. Not. R. Astron. Soc} {\bf 403}, 926 (2010).
\item R. Blandford, C. McKee, {\it Phys. Fluids} {\bf 19}, 1130 (1976).
\item T. Piran, E. Nakar, {\it Astrophys. J.} {\bf 718}, L63 (2010).
\item Y.-Z. Fan {\it et al.}, http://arxiv.org/abs/1305.1261 (2013).
\item T. Laskar {\it et al.}, http://arxiv.org/abs/1305.2453 (2013).
\item D. Giannios, {\it Astron. \& Astrophys.} {\bf 480}, 305 (2008).
\item C. Dermer {\it et al.}, {\it Astrophys. J.} {\bf 537}, 785 (2000).
\item R. Sari, A. Esin, {\it Astrophys. J.} {\bf 548}, 787 (2001).
\item B. Zhang, P. M{\'e}sz{\'a}ros, {\it Astrophys. J.} {\bf 559}, 110 (2001).
\item R.-Y. Liu {\it et al.}, {\it Astrophys. J.} {\bf 773}, L20 (2013).
\item C. Kouveliotou {\it et al.}, in prep (2013).
\item C. Dermer, A. Atoyan, {\it New J. Phys.} {\bf 8}, 122 (2006).
\item E. Blaufuss, {\it GCN Circ. 14520} (2013).
\item S. Razzaque {\it et al.}, {\it Astrophys. J.} {\bf 613}, 1072 (2004).
\item K. Murase, {\it Astrophys. J.} {\bf 745}, L16 (2012).
\end{enumerate}

\subsection*{Acknowledgments}
The Fermi LAT Collaboration acknowledges support from a number of agencies and institutes for both development and the operation of the LAT as well as scientific data analysis. These include NASA and DOE in the United States, CEA/Irfu and IN2P3/CNRS in France, ASI and INFN in Italy, MEXT, KEK, and JAXA in Japan, and the K.~A.~Wallenberg Foundation, the Swedish Research Council and the National Space Board in Sweden. Additional support from INAF in Italy and CNES in France for science analysis during the operations phase is also gratefully acknowledged.

The Fermi GBM collaboration acknowledges support for GBM development, operations and data analysis from NASA in the US and BMWi/DLR in Germany.

\clearpage

\begin{figure}
\centering
\includegraphics[scale=0.6]{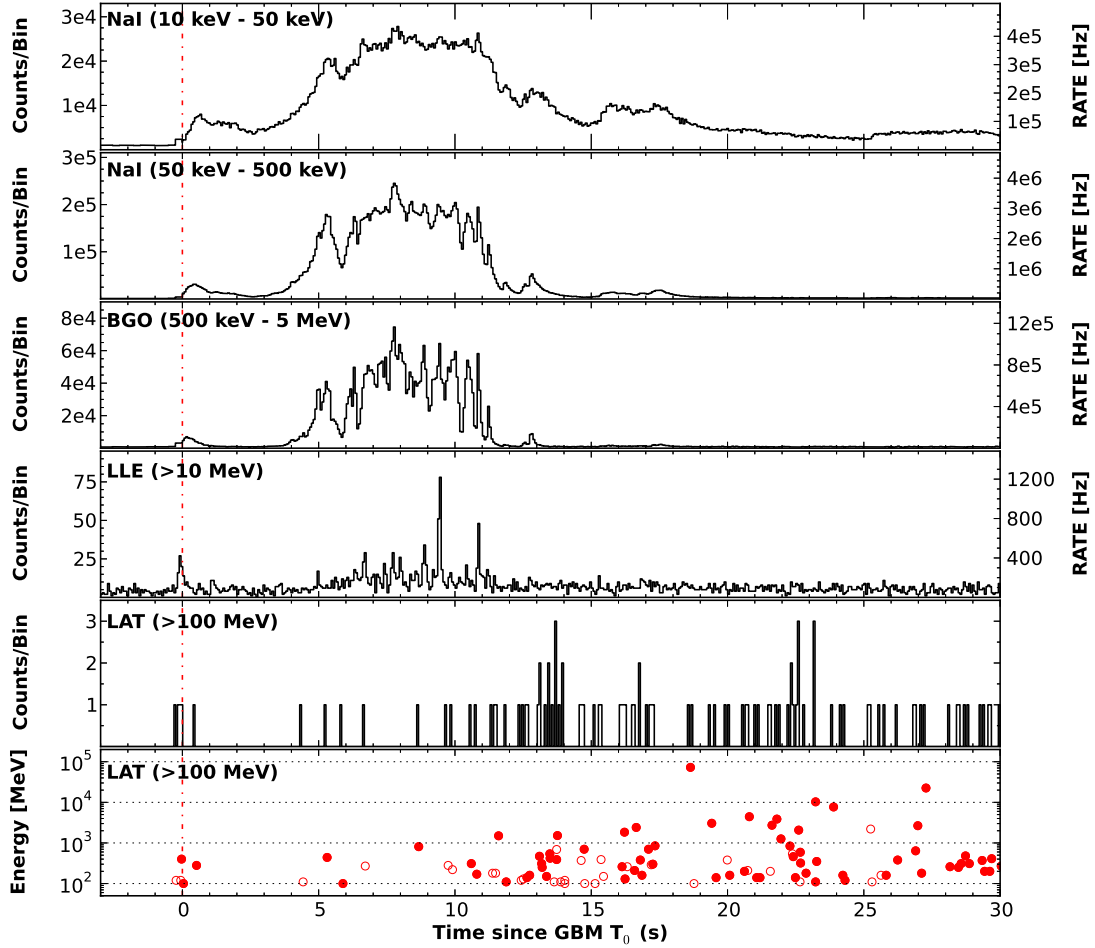}
\caption{Light curves for the Fermi-GBM and LAT detectors during the brightest part of the emission in 0.064-s bins, divided into five energy ranges. The NaI and BGO light curves were created from a type of GBM data (Continuous Time, or CTIME) that does not suffer from saturation effects induced by the extreme brightness of this GRB (SOM); for these light curves, we used NaI detectors 6, 9, and 10, and BGO detector 1. The open circles in the bottom panel represent the individual LAT $\gamma$ \emph{Transient} class photons and their energies, and the filled circles indicate photons with a $>$0.9 probability of being associated with this burst (SOM).}
\end{figure}

\begin{figure}
\center
\includegraphics[scale=0.35]{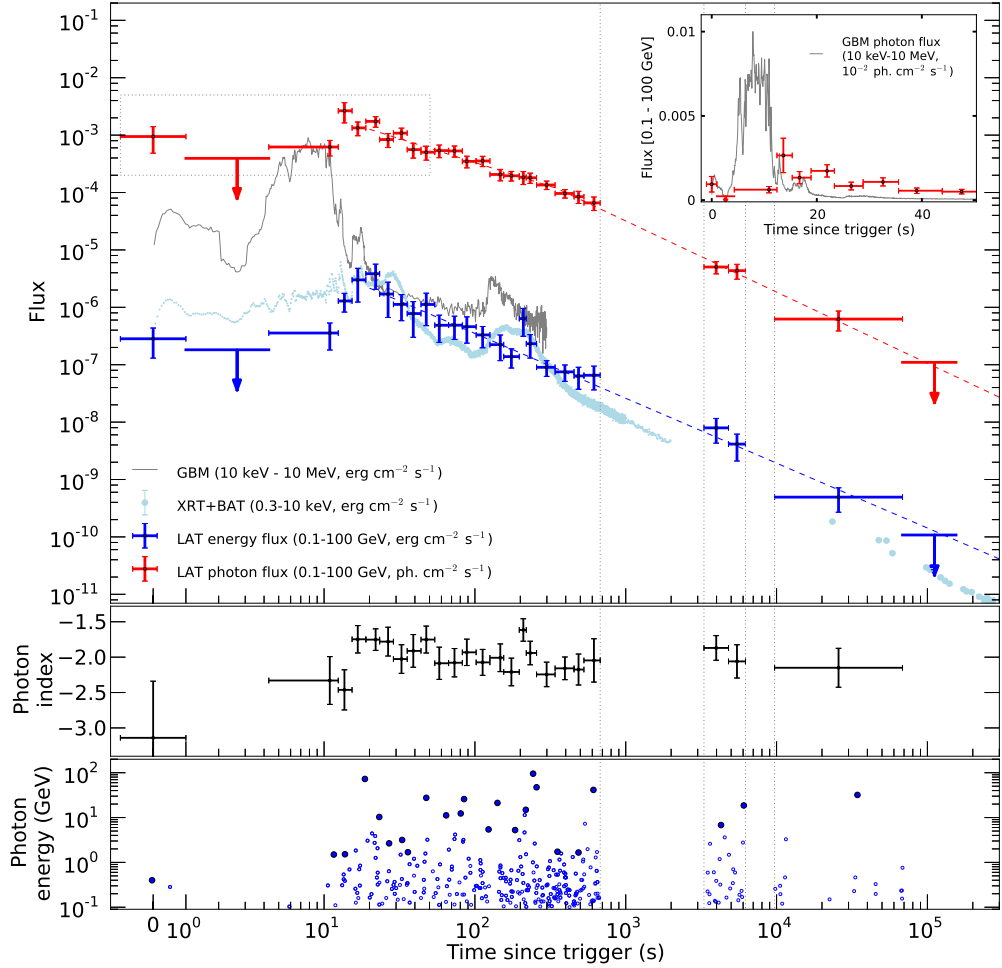}
\caption{Temporally extended LAT emission. Top: LAT energy flux (blue) and photon flux (red) light curves. The photon flux light curve shows a significant break at a few hundred seconds (red dashed line), while the energy flux light curve is well described by a single power law (blue dashed line). The 10~keV to 10~MeV (GBM, gray) and 0.3 to 10~keV (XRT+BAT, light blue) energy flux light curves are overplotted. The inset shows an expanded view of the first 50 seconds with a linear axes, with the photon flux light curve from the GBM (in units of $10^{-2}$ ph cm$^{-2}$ s$^{-1}$) plotted in gray for comparison. Middle: LAT photon index. Bottom: Energies of all the photons with probabilities $>$90\% of being associated with the GRB (SOM). Filled circles correspond to the photon with the highest energy for each time interval. Note that the photons plotted here are \emph{Source} class photons, whereas the photons in Figs. 1 and 3 are \emph{Transient} class photons (SOM). The vertical gray lines indicate the first two time intervals during which the burst was occulted by the Earth. As the ARR moved the center of the LAT FoV toward the GRB position, the effective collecting area in that direction increased, so that after $\sim\!100$~s the rate of photons increased even though the intrinsic flux decreased.}
\end{figure}

\begin{figure}
\centering
  \includegraphics[scale=0.45]{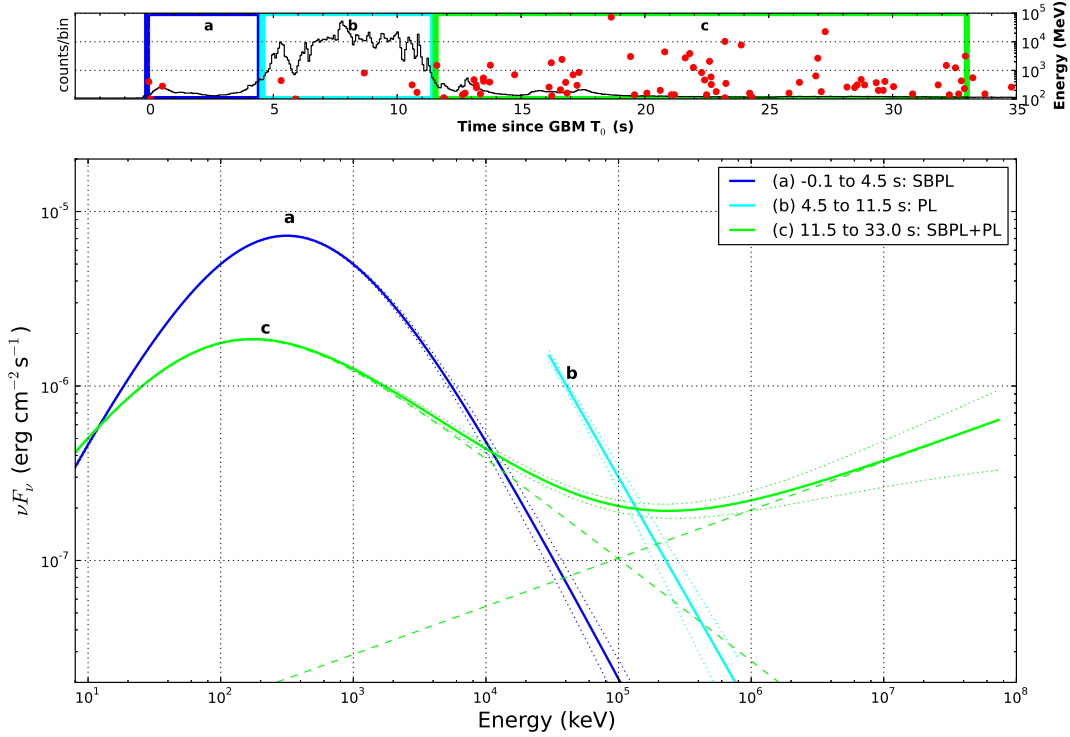}
\caption{Time-resolved spectral models for the GBM- and LAT-detected emission. Top: The combined NaI and BGO light curve from Fig.~1 (arbitrarily scaled) with the LAT-detected photons overplotted (same as the filled circles in Fig.~1). The time intervals are colored to correspond with the spectral models in the lower plot. The GBM data between 4.5 and 11.5 s are not included because they are substantially affected by pulse pileup (SOM). Bottom: The models (thick lines) that best fit the data are plotted with 1-$\sigma$ error contours (thin dashed lines). Each curve ends at the energy of the highest energy LAT photon detected within that time interval. An extra power law is statistically significant when fitting the data from $T_0 + 11.5$~s to $T_0 + 33.0$~s (SOM).}
\end{figure}

\begin{figure}
\center
\includegraphics[scale=0.35]{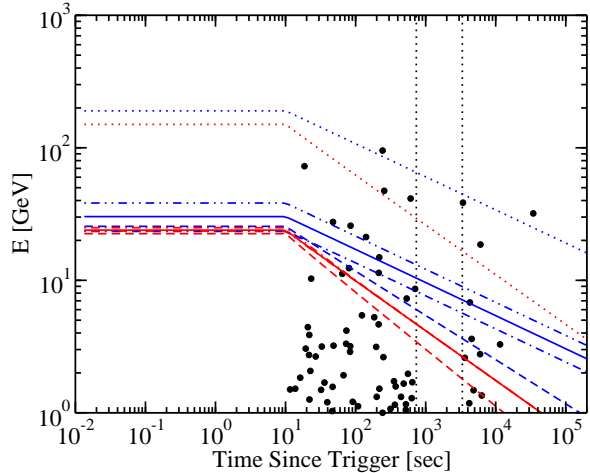}
\caption{Curves of maximum synchrotron photon energy. The black dots show the LAT detection times of photons with energies $>$1~GeV and $>\!\!90$\% probability of association with GRB 130427A. Adiabatic and radiative predictions for maximum synchrotron photon energy in uniform interstellar medium (ISM) and wind environments are plotted using the relations described in the SOM. Red and blue curves refer to the ISM and wind cases, respectively. The solid and dashed lines refer to the adiabatic and radiative cases with  $\Gamma_0 = 1000$, and the dot-dashed and double dot-dashed lines represent the adiabatic case with $\Gamma_0 = 500$ and $\Gamma_0 = 2000$, respectively. The dotted lines show an extreme possibility where acceleration takes place on the inverse of the Larmor angular frequency, in the case of an adiabatic blast wave with $\Gamma_0 = 1000$. For cases with uniform external medium, ${\cal E}_{iso}$(10$^{55}$ erg)/$n_0$(cm$^{-3}$) = 1.  The wind normalization was chosen to give the same  value of $t_{d}$ for both wind and ISM cases. The vertical dotted lines show periods of Earth avoidance when the LAT could not observe GRB 130427A.}
\end{figure}

\clearpage

\section*{Supplementary Materials}




\subsection*{Systematics}
The high flux during the brightest part of the burst saturated the GBM TTE data type for all the GBM detectors, and as a result the TTE data from $T_0 + 4.9$~s to $T_0 + 11.4$~s were unsuitable for analysis. The temporally binned CSPEC (1.024-s bins for 600 seconds after a trigger, 128 energy channels) and CTIME (0.064-s bins for 600 seconds after a trigger, 8 energy channels) data types were unaffected by saturation. However, due to the high event rate, all data types were affected by pulse pileup during the brightest part of the burst, which significantly distorts the spectrum when the detector event rate surpasses $\sim$80,000 counts per second. We excluded the data between $T_0 + 4.5$~s and $T_0 + 11.5$~s from our spectral fits of the combined GBM and LAT emission.


\subsection*{Minimum variability analysis}
The temporal variability of a GRB can give insight into the dynamics of the emission region, with direct implications for the opacity of the emitting region and, therefore, the bulk Lorentz factor. To study the variability of the $\gamma$-ray emission, we applied a modified version of the wavelet transform analysis presented in MacLachlan et al. 2013 (\emph{38}), where we used Maximum Overlap Discrete Wavelet Transform (MODWT) instead of the Discrete Wavelet Transform (DWT). The
former has superior statistical properties, especially in the context of the variance analysis (\emph{39}). We applied this technique, which does not require to assume a pulse shape, to the LLE $>$10 MeV data between $T_0+3$ and $T_0+20$~s. The minimum variability time scale measured with this technique has previously been shown to be a good proxy for the minimum rise time in GBM GRBs (\emph{40}). 

For GRB 130427A, we find an upper limit to the minimum variability timescale of $0.04 \pm 0.01$~s, so that the shortest-duration significant features in the light curve have widths of $0.08 \pm 0.02$~s. We double checked this result by using the Bayesian Block algorithm of Scargle (1998) (\emph{41}), and again found time structures with a width of $\sim$0.08~s at the $>\!\!5\sigma$ level.
Given that the LLE light curve is brightest between $T_0+5$~s and $T_0+15$~s, it follows that the minimum variability time scale we calculated is most closely associated with this time interval. However, the minimum variability time scale is a measure of the typical size of the emitting region, which is unlikely to change quickly within the prompt emission.

\subsection*{Correlated variability analysis}
We studied the temporal behavior of this GRB and searched for correlated variability between the keV and MeV-GeV emission, which, if present, would suggest that the low- and high-energy emission might arise from the same emission episode. We applied the discrete correlation function method (\emph{42}) to compare the global temporal behavior of low and high energy emission, between $T_0-5$~s and $T_0+20$~s. For the first peak at $T_0$ we found that the GBM emission lags the LLE emission, with an increasing lag with decreasing GBM energy; this is discussed in greater detail in the accompanying GBM paper on this burst (\emph{43}).

The saturation during the brightest part of this burst ($T_0 + 4.9$~s to $T_0 + 11.4$~s) prevented us from using TTE data, and as the analysis is very sensitive to bin size, the binned GBM data were not appropriate. We instead performed the analysis with the INTEGRAL/SPI-ACS light curve ($\sim$80 keV -- $\sim$10 MeV (\emph{44}) and the LLE $>$10~MeV light curve. (The number of observed photons $>$100~MeV does not provide adequate statistical power to perform this kind of timing analysis on the LAT $>\!100$~MeV data.) We found some indications that the correlation between the SPI-ACS and LLE emission decreases with increasing LLE energy, although this could be an effect of decreasing statistics.

\subsection*{Spectral analysis}
In order to further explore the relation between the low- and high-energy emission, we performed a time-resolved spectral analysis. 

\subsubsection*{Production and availability of the spectra}
We produced the spectra for LAT data by using the standard software package \emph{Fermi} ScienceTools (v9r31p1)\footnote{{\tt http://fermi.gsfc.nasa.gov/ssc/}}. LAT data above 100 MeV are available at the Fermi Science Support Center\footnote{{\tt http://fermi.gsfc.nasa.gov/ssc/data/}}, while LLE and GBM data are available in the standard Browse interface of the HEASARC\footnote{{\tt http://heasarc.gsfc.nasa.gov/W3Browse/fermi/fermille.html}}\footnote{{\tt http://heasarc.gsfc.nasa.gov/W3Browse/fermi/fermigbrst.html}}.

The procedure to obtain the observed spectra, the background spectra and the responses are described in details in (\emph{13}).

In order to make it easier to reproduce our results we present the LAT spectra and responses for all intervals for which we performed a spectral analysis in this paper (see Table S2), and for the intervals used in (\emph{7}). These files can be downloaded from our public server\footnote{{\tt https://www-glast.stanford.edu/pub\_data/}}, and they can be loaded in XSPEC\footnote{\tt{https://heasarc.gsfc.nasa.gov/xanadu/xspec}}.

\subsubsection*{Spectral modeling}
For the higher-energy emission, we used the LLE $>$30~MeV events (the energy resolution is poor between $10$ and $30$~MeV) and LAT $>$100 MeV Pass 7 \emph{Transient} class events (\emph{1}). For the GBM band, we used the TTE data between 8 keV and 900 keV from NaI detectors 6, 9, and 10 (which had the smallest source angles and therefore the best exposures to the source), and the corresponding BGO detector 1 (which is positioned on the same side of the spacecraft as NaI detectors 6, 9, and 10). For the NaI detectors, we excluded the data from the energy bins near the  K-edge (\emph{16}). The spectral fits were performed using both RMFIT (version 4.3BA) and XSPEC (version 12.8), with consistent results. We report here the fits obtained with RMFIT by minimizing the Castor statistic (CSTAT), which modifies the Cash statistic for Poisson distributed data by a data-dependent quantity so that the statistic asymptotes to $\chi^2$ as the number of counts becomes large\footnote{\tt{https://heasarc.gsfc.nasa.gov/xanadu/xspec/manual/XSappendixStatistics.html}}.


The models we considered were a power law (``PL''), a Band function (``Band''), and a smoothly broken power law (``SBPL''). The PL model has the form
\begin{align*}
f_{PL}(E) = A \left(\frac{E}{E_\textup{piv}}\right)^{\lambda},
\end{align*}
where $A$ is the normalization (in photons~cm$^{-2}$~s$^{-1}$~keV$^{-1}$) and $E_\textup{piv}$, the pivot energy (which normalizes the model to the energy range being studied), is fixed to 100~keV. In our analyses, the PL model was only used as an extra component.

The Band function is an SBPL whose curvature is defined by the spectral indices (\emph{15}). It has the form
\begin{align*}
f_{Band}(E)=
\begin{cases}
 A\left(\frac{E}{E_\textup{piv}}\right)^{\alpha} \exp\left[-\frac{(\alpha+2)E}{E_\textup{peak}}\right], & E\le\frac{\alpha-\beta}{2+\alpha}E_\textup{peak} \\
 A\left(\frac{E}{E_\textup{piv}}\right)^{\beta} \exp(\beta-\alpha) \left[\frac{\alpha-\beta}{2+\alpha}\frac{E_\textup{peak}}{E_\textup{piv}}\right]^{\alpha-\beta}, & E>\frac{\alpha-\beta}{2+\alpha} E_\textup{peak}\\
\end{cases}
\end{align*}
where $\alpha$ and $\beta$ are the low- and high-energy indices, respectively, and $E_\textup{peak}$ is the energy at which the $\nu F_\nu$ spectrum reaches a maximum. We set $E_\textup{piv} = 100$~keV.

We also used the more general SBPL model, parameterized as
\begin{align*}
f_\textup{SBPL}(E) &= A \left(\frac{E}{E_\textup{piv}}\right)^b 10^{(a-a_\textup{piv})},\text{ where} \\
a &= m\Delta\ln\left(\frac{e^q+e^{-q}}{2}\right), \quad a_\textup{piv} = m\Delta\ln\left(\frac{e^{q_\textup{piv}}+e^{-q_\textup{piv}}}{2}\right), \\
q &= \frac{\log(E/E_b)}{\Delta}, \quad q_\textup{piv} = \frac{\log(E_\textup{piv}/E_b)}{\Delta}, \\
m &= \frac{\lambda_2 - \lambda_1}{2}, \quad b = \frac{\lambda_2 + \lambda_1}{2},
\end{align*}
where $E_b$ is the break energy (in keV), $\lambda_1$ and $\lambda_2$ are the low- and high-energy indices, respectively, and $\Delta$ is the break scale (in decades of energy) (\emph{16}).

\subsection*{Temporally extended emission}
For the analysis of photons with energies $>$100 MeV arriving before the first occultation ($T_0$ + 710 s), we used the \emph{Transient} class event selection selecting events within a  $12^\circ$-radius region of interest (ROI) centered on the burst location. This region of interest was reduced to $8^\circ$ for the last interval before the Earth occultation in order to reduce background contamination by $\gamma$ rays from the Earth's limb produced by interactions of cosmic rays with the upper atmosphere. After the first occultation we used the \emph{Source} class event selection, which has more stringent background rejection cuts than \emph{Transient} class events and is better suited for analyses of long time intervals and dimmer sources. We adopted also a smaller $10^\circ$-radius ROI, which reflects the better PSF of the \emph{Source} class.

We used an unbinned maximum likelihood analysis to model the high-energy emission from the GRB. This was done with the standard software package \emph{Fermi} ScienceTools (v9r31p1)\footnote{{\tt http://fermi.gsfc.nasa.gov/ssc/}}, and in particular the Python Likelihood analysis\footnote{{\tt http://fermi.gsfc.nasa.gov/ssc/data/analysis/scitools/python\_tutorial.html}}.
 The GRB was modeled as a point source at the best available position of the afterglow (\emph{45}). For every time interval shown in Figure~2 in the main text, we tried both a power-law and a broken power-law model for the spectrum of the source, but we found no statistically significant evidence for a break. We also tried larger time intervals, and did not find significant improvements by using the broken power law instead of a simple power law. Therefore, contrary to Tam et al. (\emph{18}), we conclude that there is no statistically significant evidence for a break in the LAT energy spectrum. Instead, we explain the excess at high energy seen by Tam et al. as an effect of spectral evolution (see next section). We therefore used a power law to model the spectrum of the source.

An isotropic background component is included in the fit. For the analysis with \emph{Transient} class, the spectral properties of the isotropic component are derived using an empirical background model (\emph{46}) that is a function of the position of the source in the sky and the position and orientation of the spacecraft in orbit. For the analysis with \emph{Source} class, instead, we used the publicy available Isotropic template\footnote{{\tt http://fermi.gsfc.nasa.gov/ssc/data/access/lat/BackgroundModels.html}}. These background models accounts for contributions from both residual charged particle backgrounds and the time-averaged celestial $\gamma$-ray emission. Their normalizations are kept free during the fit. A Galactic component, based on the publicy available model for the interstellar diffuse $\gamma$-ray emission from the Milky Way, is also included in the fit, with a fixed normalization. For all time intervals, we also included all the 2FGL sources within the ROI (\emph{47}).

We usually use \emph{Source} class events for analyses of time intervals longer than 100~s. However, for GRB 130427A, the signal-to-noise was in \emph{Transient} than \emph{Source} class photons up until the first Earth occultation. We obtained consistent results with \emph{Source} class events, although the statistical uncertainties were larger.

\subsection*{Spectral Energy Distributions}
We evaluated the spectral energy distribution (SED) from the LAT data for several different time intervals, using the following procedure: In each time bin, we first performed an unbinned maximum likelihood analysis selecting the full energy range (100 MeV to 100 GeV) using Pass7 v6 \emph{Source} class photons. We selected events within $10^\circ$ from the location of GRB 130427A. The diffuse Galactic background component normalization was fixed, while the normalization of the isotropic background component was left free. We modeled GRB 130427A as a point source with either a spectrum described by a simple power law with normalization and index free or a spectrum described by a broken power laws. We recorded the value of the normalization of the isotropic component for the best fit model.

Then, we selected events for six energy ranges (100 MeV--237 MeV, 237 MeV--562 MeV, 562 MeV--1.3 GeV, 1.3 GeV--3.2 GeV, 3.2 GeV--11.6 GeV, 11.6 GeV--100 GeV) and we independently performed a likelihood analysis in each energy bin, fixing the isotropic background component normalization to the value previously obtained. The normalization and the photon index of the GRB source were left free. 
We calculated the SED points by multiplying the integrated flux in each energy bin by the square of the average energy calculated using the best-fit power-law index. We report our results in Figure~S1, where the vertical bar position coincides with the estimated average energy.  The choice of the time bins was driven by features in the LAT light curve of Figure~2 (intervals \emph{a}, \emph{b}, and \emph{d}) and the choice in (\emph{18}) (intervals \emph{c}, \emph{e}).

\begin{figure}[htbp]
   \centering
\includegraphics[scale=0.3]{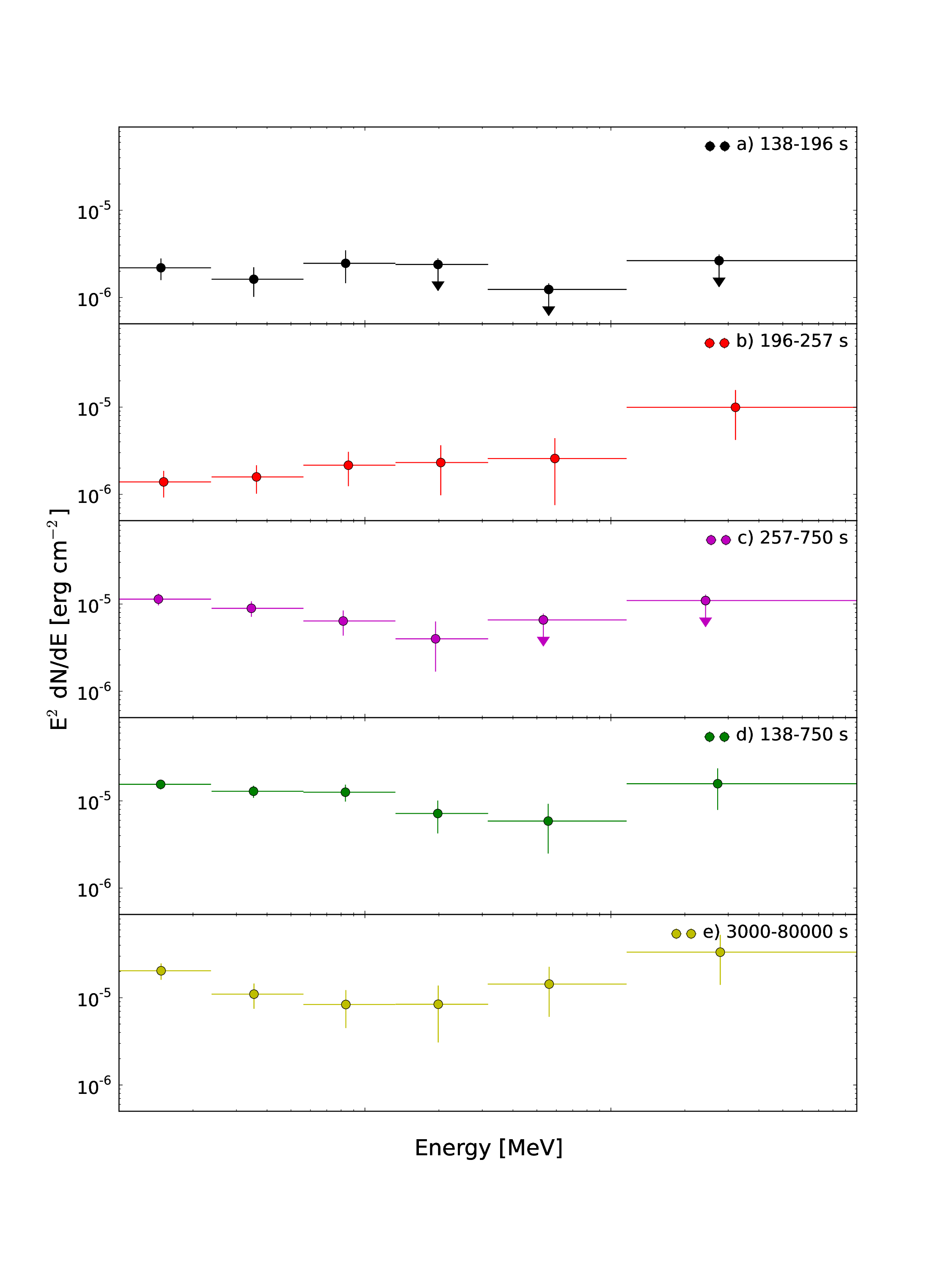}
   \caption*{Figure S1. Spectral Energy Distribution of LAT data in five different time intervals (a=138--196 s, b=196--257 s, c=257--750s, d=138--750 s, e= 3000s--80000~s).}
   \label{fig:SED}
\end{figure}

We compared our results with Tam et al. (\emph{18}), and we found that in all the time intervals, the simple power-law model better describes the data well, and the broken power law is not statistically required.
Also, we notice that the SED in interval \emph{d} is consistent with the sum of the hard component spectrum of interval \emph{b} and  the soft component spectrum in interval \emph{a} and \emph{c}. This points to the spectral evolution as source of the high-energy excess, rather than the simultaneous co-existence of two components, as claimed by (\emph{18}).

\subsection*{Constraints on $\Gamma$}

A minimum bulk Lorentz factor $\Gamma_{min}$ for GRB 130427A can be calculated by using the variability timescale $t_{var}$, obtained as described above, to estimate the size scale $\Delta r^\prime \cong \Gamma c t_{var}/(1+z)$  of the emission region. The comoving photon energy density $u^\prime(\nu^\prime)$ in the fluid frame is related to the measured spectral luminosity through the expression $\nu L_\nu = 4\pi d_L^2 \Gamma^2 u^\prime(\nu^\prime)$, where $\nu^\prime \cong \nu/\Gamma$ (for details, see SOM in (\emph{48})). The value for $\Gamma_{min}\cong 500$ is derived for the 73~GeV photon detected at 19.06 s after $T_0$, as described in the text. This uses a value of $t_{var}$ determined from the LLE light curve (see Minimum Variability Analysis). The uncertainties in $\Gamma_{min}$ quoted in the text are formal uncertainties, related to the uncertainty in $t_{var}$, and do not include the uncertainties related to the errors in the best fit parameters and photon statistics, the latter giving $\sim 20$\% uncertainty of $\Gamma_{min}$.

The initial bulk Lorentz factor $\Gamma_0$ can be determined by assuming that the duration of the brightest LAT emission corresponds to the deceleration timescale $t_{d}$.  For a uniform circumburst medium with density $n = n_0($cm$^{-3})$, $t_{d} = (3{\cal E}_{iso}/4\pi \Gamma_0^8 n m_pc^2)^{1/3}/c \equiv 10t_1$~s, implying $\Gamma_0 \cong 540 ({\cal E}_{55}/n t_1^3)^{1/8}$,
where ${\cal E}_{iso}\equiv 10^{55}{\cal E}_{55}$~erg, which is larger than  ${\cal E}_{\gamma,iso}$ by a factor of $\approx 7$. Relations for determining $\Gamma_{min}$ in a medium with a power-law density gradient can be found in (\emph{49}).

\subsection*{Maximum synchrotron energy}

In the coasting and deceleration phases, the evolution of $\Gamma$ with radius $r$ for an adiabatic blast wave in a uniform circumburst medium (the ISM case) can be parameterized by the function $\Gamma(r) \cong {\Gamma_0/ \sqrt{1+(r/r_{d})^3}}$. Generalization to different radiative regimes and circumburst density gradients can be made through the expression
\begin{equation}
\Gamma(r) = \frac{\Gamma_0}{\sqrt{1+a_qx^q}}\equiv \Gamma_0 G_q(x)\;,\;x\equiv \frac{r}{r_d}\;,
\label{Gammar}
\end{equation}
where $r_d = (3{\cal E}_{iso}/4\pi n m_pc^2 \Gamma_0^2)^{1/3}$.
By choosing $q$ and $a_q$ appropriately, these  parameterizations recover 
 the exact asymptotes in the deceleration phase. For the adiabatic 
Blandford-McKee (1976) solution in the ISM case, $q = 3$ and $a_3 = 0.70$, 
which is obtained from the expression 
$r = (17{\cal E_{iso}}/16\pi \Gamma^2 n m_pc^2)^{1/3}$ (\emph{24, 50}). Alternately, one can take $a_q = 1$ and define $r_d$ such that $\Gamma(r)$ recovers the self-similar solution.

The maximum synchrotron photon energy in the stationary frame of the 
black hole is
$\epsilon_{max,syn} =  \delta_{\rm D} \epsilon^\prime_{cl}\,$
(in units of $m_e c^2$), where the Doppler factor 
$\delta_{\rm D} = [\Gamma(1-\beta\mu)]^{-1} \rightarrow  {2\Gamma/ (1+\Gamma^2\theta^2)}$
in the limit $\Gamma(r)\gg 1$ and $\theta\ll 1$. Equating the inverse of the timescale to 
execute a complete Larmor orbit with the timescale for synchrotron losses gives 
the classical, radiation-reaction limited value of 
$\epsilon^\prime_{cl} = {27/ 16\pi\alpha_f}\cong 74 \cong 38/0.511$, 
where $\alpha_f = 1/137$ is the fine-structure constant. A more stringent but less physically
plausible criterion equates the synchrotron timescale with the inverse of the Larmor angular frequency,
which will allow $\epsilon_{cl}^\prime$ to be a factor of $2\pi$ larger. 

The time $t_*$ in the stationary frame relates to the time $t$ in the observer frame for emission at radius $r$ and angle $\theta = \cos^{-1}\mu$ to the line of sight through the expression $t = t_* - r\mu/c$. Taking $r$ as the radius of the forward shock, $dr = \beta_{sh} c dt_* = \beta_{sh}\Gamma_{sh} c dt^\prime = \beta_{sh} \delta_{{\rm D},sh}\Gamma_{sh} c dt$, so that
\begin{equation}
dt = \frac{dr}{\beta_{sh} \Gamma_{sh}\delta_{{\rm D},sh}c} \rightarrow \frac{dr}{2\Gamma_{sh}^{2}c} (1+\Gamma_{sh}^2\theta^2 )\cong \frac{dr}{4\Gamma^2c} (1+2\Gamma^2\theta^2)
\label{dt}
\end{equation}
in the limit $\Gamma\gg 1$ and $\theta \ll 1$, where the shocked fluid Lorentz factor $\Gamma \cong \Gamma_{sh}/\sqrt{2}\gg 1$.
This differential equation can be rewritten in terms of $r_d$ and $t_d= r_d/2\Gamma_0^2 c$ using the variables $\tau = t/t_d$, and $N = \Gamma_0\theta$, giving $2d\tau = dx(1+2 N^2 +a_q x^q)$, with solution
$\tau = [(1+2N^2)x+a_q x^{q+1}/(q+1)]/2$.
This expression was numerically inverted to give $x(\tau,N)$.
The maximum synchrotron photon energy $\epsilon_{max,syn}$ at time $t = t_d\tau$
is found by numerically scanning through $N=\Gamma_0\theta$ in the expression
\begin{equation}
\epsilon(\tau,N) = \frac{2\Gamma_0 G_q[x(\tau,N)]\epsilon^\prime_{cl}}{1+N^2 G_q^2[x(\tau,N)]}\;
\label{tau}
\end{equation}
to find its maximum value.

When $t_{dec}>t_d$, that is, when the engine duration $T_{GRB}>t_d$, the above computation is modified somewhat. When $t > t_{dec}$, the computation is the same as for the case $t_d<T_{GRB}$. When  $t < t_{dec}$, we let $\Gamma(t) \rightarrow \Gamma_0(t_{dec})$, where
$\Gamma_0(t_{dec}) = (2ct_{dec})^{-3/8}(3{\cal E}_{iso}/4\pi n m_pc^2)^{1/8}\cong 1280 [t_{dec}({\rm s})]^{-3/8}
({\cal E}_{55}/n)^{1/8}$. 
Note, however, that even for $T_{GRB} < t_{d} = t_{dec}$, the shell spreads radially and the reverse shock becomes mildly relativistic at $ t \sim t_{dec}$. Thus $\Gamma(t_{dec})$, representing the bulk Lorentz factor of the newly shocked material around that time,
could be a factor $\approx 2$ lower than given by this expression.
Here we use the more optimistic expression for $E_{syn,max}$.

\begin{figure}
\center
\includegraphics[width=14.0cm]{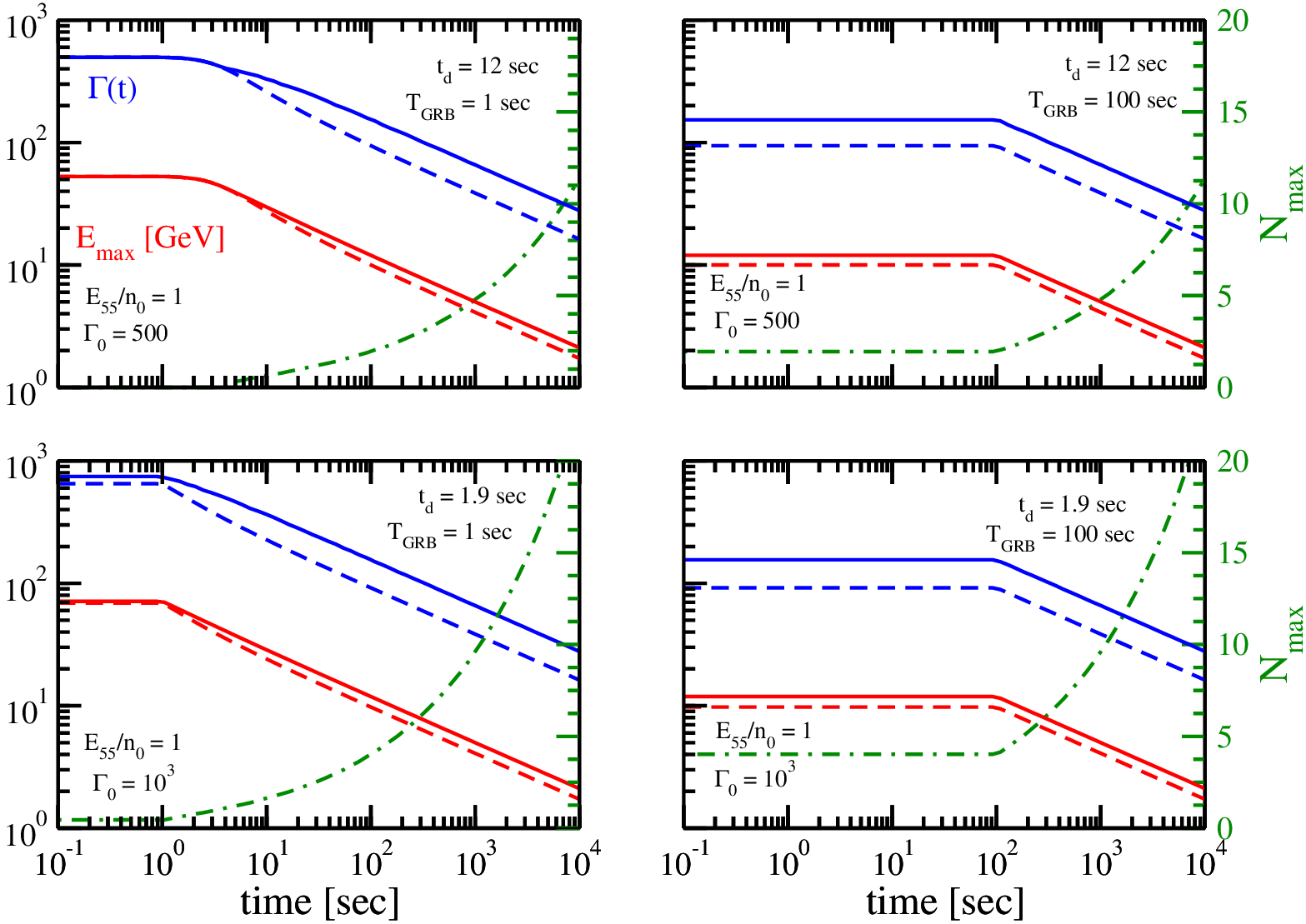}
\caption*{Figure S2. Adiabatic evolution of the bulk Lorentz factor $\Gamma$ with time $t$ in a uniform circumburst medium, showing the on-axis bulk Lorentz factor $\Gamma (t,N=0)$ (solid blue curves), and the value of $\Gamma(t,N_{max})$ that produces the most energetic synchrotron 
photon energy at time $t$ (dotted blue curves), where $N\equiv \Gamma_0\theta$ and $N_{max} = \Gamma_0 \theta_{max}$. In all cases, ${\cal E}_{55}/n_0 = 1$. The maximum synchrotron photon energy from on-axis emission is given by the solid red curves, and from off-axis emission by the dashed red curves. At early times, the two are equal, and given by $2\sqrt{2}\times 38$ MeV $\times\,\Gamma_0/1.34 \cong$ 40 GeV for $\Gamma_0 = 500$, $T_{GRB}\ll t_{dec}$. The green dot-dashed curve gives the value of $N_{max}$ that maximizes the maximum synchrotron photon energy. (a) Top left: $\Gamma_0$ = 500, $T_{GRB} = 1$ s. (b) Top right:  $\Gamma_0$ = 500, $T_{GRB} = 100$s. (c) Bottom left:  $\Gamma_0$ = 1000, $T_{GRB} = 1$ s. (d) Bottom right:  $\Gamma_0$ = 1000, $T_{GRB} = 100$ s.}
\label{0229}
\end{figure}

Figure~S2 shows results for an adiabatic blast wave in the ISM case with $\Gamma_0 = 500$ and 1000, and $T_{GRB}=1$ and 100 s, for $z=0.34$ and ${\cal E}_{55}/n_0 = 1$. The cases with $T_{GRB}=1$ and 100 s are illustrative, showing the effects of injection when $T_{GRB} < t_{dec}$ (the impulsive regime), and when $T_{GRB} > t_{dec}$ (the extended-engine regime). We use $T_{GRB} = 10$ s when comparing with data from GRB 130427A, as this value corresponds to the timescale during which most of the energy is emitted,  as shown by the GBM light  curves. Note that $N_{max}$ is nonzero during the coasting phase when $T_{GRB}>t_d$.

The curves, as labeled, show the on-axis values of $\Gamma(t,N=0)$, 
the off-axis values of $\Gamma(t,N=N_{max})$ that produce the highest energy 
synchrotron photons. Also shown are the on-axis and off-axis values
of $E_{max,syn}$. These values are nearly identical because the larger 
Lorentz factors of the off-axis emission compete with reduction of Doppler 
boosting when the emitting plasma is moving at an angle to the line of sight.
The angle where the highest 
energy synchrotron photon originates, in units of $1/\Gamma_0$, is shown by the long dashed curve. 
The results of Figure~S2 agree  with Eq.\ (4) of (\emph{25}), scaled by the factor
$27/16\pi$.

\begin{table}
\centering\tiny
\begin{tabular}{l|l|r@{ $\pm$ }lr@{ $\pm$ }lr@{ $\pm$ }lr@{ $\pm$ }lr@{ $\pm$ }lr@{ / }l}
  \hline
  \textbf{Time interval}  & \multicolumn{1}{c|}{\textbf{Model}}
                 & \multicolumn{2}{c}{Norm.}
                 & \multicolumn{2}{c}{$E_{break}$}
                 & \multicolumn{2}{c}{$\Delta$}
                 & \multicolumn{2}{c}{$\lambda_1$}
                 & \multicolumn{2}{c}{$\lambda_2$}
                 & \multicolumn{2}{c}{C-STAT / dof}
  \\ \hline \hline
  (a) -0.1 to    & SBPL
                 &  0.3145 & 0.0015
                 &     291 & 23
                 & \multicolumn{2}{c}{1.08}
                 &  -0.334 & 0.025
                 &  -3.394 & 0.072
                 &     700 & 474
  \\
   \quad 4.5 s  & 
                 & \multicolumn{2}{c}{}
                 & \multicolumn{2}{c}{}
                 & \multicolumn{2}{c}{}
                 & \multicolumn{2}{c}{}
                 & \multicolumn{2}{c}{}
                 & \multicolumn{2}{c}{}
  \\ \hline
  (b) 4.5 to    &  PL
                 &   189 & 165
                 & \multicolumn{2}{c}{--}
                 & \multicolumn{2}{c}{--}
                 &  -3.34 & 0.15
                 & \multicolumn{2}{c}{--}
                 &     20 & 15
  \\
  \quad 11.5 s   & 
                 & \multicolumn{2}{c}{}
                 & \multicolumn{2}{c}{}
                 & \multicolumn{2}{c}{}
                 & \multicolumn{2}{c}{}
                 & \multicolumn{2}{c}{}
                 & \multicolumn{2}{c}{}
  \\ \hline
  (c) 11.5 to    & SBPL
                 &  0.1000 & 0.0003
                 &    39.1 & 3.3
                 & \multicolumn{2}{c}{1.11}
                 &  -0.623 & 0.050
                 &  -2.408 & 0.012
                 &    1028 & 474
  \\
  \quad  33.0 s  & 
                 & \multicolumn{2}{c}{}
                 & \multicolumn{2}{c}{}
                 & \multicolumn{2}{c}{}
                 & \multicolumn{2}{c}{}
                 & \multicolumn{2}{c}{}
                 & \multicolumn{2}{c}{}
  \\ \cline{2-14}
                 & SBPL
                 &  0.1003 & 0.0005
                 &    60.0 & 7.8
                 & \multicolumn{2}{c}{1.11}
                 &  -0.784 & 0.046
                 &  -2.515 & 0.037
                 &     984 & 472
  \\
                 & \,+PL
                 & \multicolumn{2}{c}{$(4.2_{-3.0}^{+6.0}) \times 10^{-4}$}
                 & \multicolumn{2}{c}{--}
                 & \multicolumn{2}{c}{--}
                 & -1.66 & 0.13
                 & \multicolumn{2}{c}{--}
                 & \multicolumn{2}{c}{}
  \\ \hline
  (d) 33.0 to   & SBPL
                 &  0.0133 & 0.0001
                 &     7.3 & 3.2
                 & \multicolumn{2}{c}{1.21}
                 &    0.37 & 0.59
                 &  -2.238 & 0.012
                 &     406 & 241
  \\
   \quad 196.0 s  & 
                 & \multicolumn{2}{c}{}
                 & \multicolumn{2}{c}{}
                 & \multicolumn{2}{c}{}
                 & \multicolumn{2}{c}{}
                 & \multicolumn{2}{c}{}
                 & \multicolumn{2}{c}{}
  \\ \cline{2-14}
                 & SBPL
                 & 0.0133 & 0.0001
                 &    7.3 & 3.6
                 & \multicolumn{2}{c}{1.21}
                 &   0.41 & 0.65
                 & -2.247 & 0.020
                 &    401 & 239
  \\
                 & \,+PL
                 & \multicolumn{2}{c}{$(1.6^{+2.8}_{-1.0})\times 10^{-6}$}
                 & \multicolumn{2}{c}{--}
                 & \multicolumn{2}{c}{--}
                 &  -1.22 & 0.68
                 & \multicolumn{2}{c}{--}
                 & \multicolumn{2}{c}{}
  \\ \hline
\end{tabular}
\caption*{Table S1. The prompt emission spectral fit parameters, using GBM, LLE, an LAT data (except for interval b, which only uses LLE and LAT data). All times are relative to $T_0$. We tried to fit a smoothly broken power law (``SBPL'') as well as an SBPL with an extra power-law (``PL'') component. Parameters are: normalization in photons cm$^{-2}$ s$^{-1}$ keV$^{-1}$, $E_{break}$ in keV, break scale $\Delta$ (for SBPL) in decades of energy, low-energy index $\lambda_1$, and high-energy index $\lambda_2$. C-STAT is the Castor statistic (a modified version of the Cash statistic). See text for definitions of models. For all time intervals that were fit with a SBPL, we fixed the break scale $\Delta$ to the best-fit value. The extra PL component is statistically significant in interval c ($\Delta$CSTAT = 44 for 2 degrees of freedom). In interval \emph{d}, we used only NaI detector 6.}
\label{spectralFits}
\end{table}

\begin{table}
\centering
\begin{tabular}{rrr@{.}l}
  \hline
      \multicolumn{1}{c}{\emph{E}}
    & \multicolumn{1}{c}{\emph{E$_\textup{rf}$}}
    & \multicolumn{2}{c}{$T-T_0$}
  \\ \hline \hline
  95  &  128  &  243&55    
  \\ \hline
  73  &  97   &  19&06     
  \\ \hline
  47  &  63   &  256&70    
  \\ \hline
  41  &  55   &  611&01   
  \\ \hline
  39  &  52   &  3410&26  
  \\ \hline
  32  &  43   &  34366&58 
  \\ \hline
  28  &  37   &  48&01    
  \\ \hline
  26  &  35   &  85&16     
  \\ \hline
  21  &  21   &  141&53   
  \\ \hline
  15  &  20   &  217&89   
  \\ \hline
\end{tabular}
\caption*{Table S2. The 10 highest-energy LAT photons with probability $>\!1 - 10^{-3}$ of being associated with the GRB as opposed to background, determined using a likelihood analysis. $E_\textup{rf}$ is the photon's rest frame energy at the redshift $z=0.34$. All photons are \emph{Source} class. $E$ and $E_\textup{rf}$ are in GeV, $T-T_0$ in s.}
\label{highestEnergies}
\end{table}

\end{document}